\documentclass[11pt,twoside]{article}

\usepackage{amsfonts}
\usepackage{amsmath}
\usepackage{amssymb}
\usepackage{color}

\newcommand{\bean}{\begin{eqnarray*}}
\newcommand{\eean}{\end{eqnarray*}}
\newcommand{\benu}{\begin{enumerate}}
\newcommand{\eenu}{\end{enumerate}}
\newcommand{\eea}{\end{eqnarray}}
\newcommand{\bea}{\begin{eqnarray}}

\newtheorem{Theorem}{Theorem}

\newtheorem{Lemma}{Lemma}

\newcommand{\be}{\begin{equation}}
\newcommand{\ee}{\end{equation}}

\newcommand{\N}{{\mathbb N}}

\newcommand{\R}{{\mathbb R}}

\newcommand{\s}{\sigma}
\newcommand{\LSA}{{\rm LSA}}
\newcommand{\LD}{{\rm LD}}
\newcommand{\Meas}{{\rm Meas}}

\def\d{\delta}
\def\l{\lambda}
\def\o{\circ}
\def\supp{{\rm supp}}

\newcommand{\ben}{\begin{enumerate}}
\newcommand{\een}{\end{enumerate}}
\newcommand{\bit}{\begin{itemize}}
\newcommand{\eit}{\end{itemize}}
\newcommand{\edoc}{\end{document}}

\newcommand{\bdefi}{\begin{definition}}
\newcommand{\btheo}{\begin{theorem}}
\newcommand{\bprop}{\begin{proposition}}
\newcommand{\brema}{\begin{remark}}
\newcommand{\bcoro}{\begin{corollary}}
\newcommand{\blemm}{\begin{lemma}}
\newcommand{\bexam}{\begin{example}}

\newcommand{\edefi}{\end{definition}}
\newcommand{\etheo}{\end{theorem}}
\newcommand{\eprop}{\end{proposition}}
\newcommand{\erema}{\end{remark}}
\newcommand{\ecoro}{\end{corollary}}
\newcommand{\elemm}{\end{lemma}}
\newcommand{\eexam}{\end{example}}

\newcommand{\V}{\noindent}
\newcommand{\ci}{\circ}
\def\qed{\ensuremath{\quad\Box\quad}}

\title{On the von Neumann rule in quantization}

\author{Olaf M\"uller\footnote{Humboldt-Universit\"at, Institut f\"ur Mathematik, Unter den Linden 6, 10099 Berlin}}

\date{\today}

\begin{document}

\maketitle

\begin{abstract}
\V We show that any linear quantization map into the space of self-adjoint operators in a Hilbert space violates the von Neumann rule on post-composition with real functions. 
\end{abstract}  

\section{Introduction and main results}

Physics today has the ambition to be entirely mathematically derivable from two fundamental theories: gravity and the standard model of particle physics. Whereas the former is a classical field theory with only mildly paradoxical features such as black holes, the latter is not so much a closed theory as rather a toolbox full of complex algorithms and ill-defined objects. Moreover, as it uses a form of canonical quantization, the non-mathematical, interpretational subtleties of quantum mechanics, such as the measurement problem, can also be found in the standard model. Despite of these problems, the standard model is very successful if used by experts inasfar as its predictions are in accordance with a large class of experiments to an unprecedented precision. Its canonical quantization uses a quantization map $Q: G \rightarrow \LSA(H)$ from some nonempty subset $G \subset C^0(C)$ of classical observables, where $C$ is the classical phase space, usually diffeomorphic to the space of solutions, and $ \LSA (H)$ is the space of linear self-adjoint (s.-a.) maps of a Hilbert space to itself.\footnote{More exactly, the image of $Q$ consists of essentially s.-a. operators with a common dense domain of definition (usually a Schwartz space), which is, due to uniqueness of the s.-a. extension, a modification without consequences for the results of this article.}

\V There is a widely accepted list of desirable properties for a quantization map  going back to Weyl, von Neumann and Dirac (\cite{hW}, \cite{jN}, \cite{pD}):

\begin{enumerate}
\item $Q$ is $\R$-linear (in particular , $G$ is a real vector space);
\item $Q$ is unit-preserving, i.e. $Q (1) = I_H $ where $I_H$ is the identity in $H$;
\item {\bf von Neumann rule}: $Q$ is invariant under postcomposition with smooth maps $\R \rightarrow \R$, i.e. for all $f \in G , \psi \in C^\infty (\R, \R)  $ we have $\psi \o f  \in G$ and $Q( \psi \ci f) = \psi ( Q(f) ) $ in the sense of functional calculus;
\item $ \exists p, q \in G \exists c \in i \R : [ Q(p), Q(q) ] = c I_H$ (canonical commutation). 
\end{enumerate}

\V The last item is weaker than the assignment used in canonical quanization\footnote{where $C$ is a space of sections of a bundle $\pi: E \rightarrow N$ whose fiber is the (co-)tangent space of a manifold with local adapted coordinates $(x_i, p_i)$, $H $ is some space of complex (polarized) functions on $C$, and for a function $u$ on $N$,  $ Q(u \cdot x_i)  $ is the operator of multiplication with $ u \cdot x_i$ whereas $ Q(u \cdot p_i) $ is the closure of $ u \cdot \partial_i$, modulo the correspondence between vector fields along a function $f$ and vectors at a function within the space of functions. Often, this assignment is first defined in the context of quantum {\em mechanics}, i.e., for $N$ being a point, and in the limit of $u$ tending to a delta distribution, and only later transferred to quantum field theory.}. A related requirement is that $Q$ is a Poisson representation in the sense that it takes the Poisson bracket to an imaginary multiple of the commutator.

\V The motivation for the Neumann rule is that {\em measuring $f$ is the same as measuring $ \psi \o f$}, and the effect of $\psi$ amounts to a mere relabelling of the scale of the measuring apparatus, if we recall that measuring a quantity simply means coupling a macroscopic quantity homeomorphically to it. If somebody changes the scale of a measurement apparatus, applying to it a map $\phi: \R \rightarrow \R$, the modified apparatus still extracts the same exact amount of information from the system, which is precisely what is encoded in the von Neumann rule, at least if $\phi$ is a homeomorphism onto its image. 

\V The Neumann property, deeply rooted in the axioms of quantum theory, follows e.g. from the Born rule (which in turn, via Gleason's theorem, follows from the probabilistic interpretation of Hilbert space geometry, where projections correspond to 'yes/no'-questions with 'and' related to the intersection, 'or' to the closed linear span, 'not' to the orthogonal complement):

\V By the Born rule (which, as a physical statement, contains the mathematically undefined term 'measurement'), the probability $p(f, \lambda, v)$ of measuring $\lambda$ for a classical observable $f$, if the system is in the state $v$, is $\langle v, P_{Q(f), \lambda} v \rangle $, where, for an operator $A$, $P_{A, \lambda}$ is the orthogonal projection onto the eigenspace $E_{A, \l} := \ker (A - \l I_H)$ of $A$ to the eigenvalue $\lambda$. As measuring $f$ is measuring $\psi \o f$, we have $p(f, \lambda,v ) = p(\psi\o f, \psi(\lambda),v) $, thus

$$ \forall v \in H: \langle v,P_{Q(f), \lambda} v \rangle = \langle v, P_{Q(\psi \o f), \psi(\lambda)} v \rangle ,$$

\V so $ P_{Q(f), \lambda}= P_{Q(\psi \o f), \psi(\lambda)} $ by polarization, and the Neumann rule follows.

\V Of course, one should additionally ask for other properties such as continuity and functoriality of $Q$ in an appropriate category. But unfortunately, already the four properties above cannot be satisfied at once by the same map. The proof of that fact goes back to Arens and Babbitt \cite{AB} and Folland \cite{Folland}, see also the excellent review article by Ali and Engli\v{s} \cite{AE}. Engli\v{s} also obtained the remarkable result \cite{mE2} that with canonical quantization as above (i.e., where $Q$ maps $x_j$ to 
the operator of multiplication with $x_j$ and $p_j$ to a multiple of the closure of $i \cdot \partial_j$), there is no Neumann map $Q$ which is a Poisson representation, without assuming linearity of $Q$ or even of $Q(f)$!

\V To the best of the author's knowledge, all quantization schemes so far try to satisfy the von Neumann property only approximately, e.g., modulo higher orders of $\hbar$. But if we assume quantum theory to be a fundamental theory, the exact validity of the Neumann rule is central, as explained above.  One could hope that it is possible to conversely satisfy the von Neumann property exactly at the expense of the canonical commutation relation, which then can be satisfied only approximately. This note shows that this kind of approach is doomed to failure. We first note that the von Neumann rule implies that the domain $G$ of $Q$ is a representation space for the monoid $\LD := \{ f \in C^\infty (\R, \R) \vert f {\rm \  is \ a \ diffeomorphism \ onto \ its \ image} \}$. Conversely, for a representation space $G$ of $\LD$ let us call a map $Q:   G \rightarrow \LSA (H)$

\begin{itemize}
\item{{\bf Neumannian} iff for all $f \in G$ and all $\psi \in \LD$ we have $Q(\psi \o f) = \psi(Q(f))$ in the sense of functional calculus;}
\item{{\bf Abelian} iff $Q(G)$ is an Abelian subalgebra of $\LSA(H)$;}
\item {{\bf local} iff $H$ is a Sobolev space of sections of a Hermitean bundle $\pi$ over a Hilbert manifold $F$ equipped with a Borel measure and $Q(f) \vert_{\Gamma_{C^\infty} (\pi)}$ does not increase supports for all $f \in G $.}
\end{itemize}

\V The motivation for the last property (locality) is that in Geometric Quantization and other quantization schemes, a guiding idea is to interpret a quantum state as a superposition of classical states, more precisely, a polarized complex probability distribution over the set of classical states, so that in this case $F=C$. This anchoring in spacetime is an aspect sometimes neglected by the abstract operator algebra formulation, but recall that there is exactly one isomorphism class of separable Hilbert spaces, thus in this case the main physical information is not in the space itself but in its identification with probability densities located in spacetime. However, this notion of locality is stronger than the spacetime notion of locality linked to functoriality of quantization as in, for example, \cite{BFV} or \cite{cF}.

\bigskip

\V The results of this article are: 

\begin{Theorem}
\label{LinearNeumann}
Every $\R$-linear Neumannian map is Abelian.
\end{Theorem}

\V (of course, $\R$-linearity of $Q$ presupposes that $G$ is a real vector space).

\V {\bf Remark.} The article \cite{AE} gives a similar statement as Theorem \ref{LinearNeumann} without proof, referring apparently to \cite{mE}, where a proof is given on the additional basis of Assumption 4 of our list above (existence of two quantum operators satisfying the canonical commutation relation).

\V As we also want to prove something on local Neumannian maps, we will need an infinite-dimensional version of Peetre's theorem, proven closely along the lines of the proof for the finite-dimensional case:

\begin{Theorem}[Peetre's Theorem for Hilbert manifolds]
\label{Peetre}
Let $M$ be a Hilbert manifold and let $\pi: E \rightarrow M$ and $\psi: F \rightarrow M$ be smooth Fr\'echet vector bundles over $M$. Let $L: \Gamma_{C^\infty} (\pi) \mapsto \Gamma_{C^\infty} (\pi)$ be a morphism of sheaves that is {\bf support-nonincreasing}, i.e. $\supp (Ls) \subset \supp (s) $ for all $s \in \Gamma_{C^\infty} (\pi)$. Then for all $p  \in M$ there is an open neighborhood $U$ of $p$ and there is $k \in \N$ such that $L \vert_U$ is a differential operator of order $k $, i.e. there is a vector bundle homomorphism $u: J^k \pi \rightarrow \psi$ with $L \vert_U = u \o j^k \o r_U$, where $r_U$ is restriction of sections to $U$.
\end{Theorem}

\begin{Theorem}
\label{LocalNeumann}
Any local (not necessarily linear) Neumannian map is Abelian.
\end{Theorem}

\V An interesting question is whether the same statement is true if one replaces the Hilbert manifold $F$ in the definition of locality with a Fr\'echet manifold.

\bigskip

\V As noncommutativity is precisely the essence of every quantum theory in the sense that the order of measurements changes the result in a statistically reliable way and taking into account the importance of the Neumann property, these theorems mean that {\em any physically valid quantization map should be neither linear nor local}. 

\bigskip

\V The next section is devoted to the proofs of the theorems. The final section draws some conclusions for fundamental physics, addressing primarily the questions whether quantization is a valid concept at all and whether nonlinear functions on phase space are physically observable. 

\V The author wants to thank Dirk Kreimer for useful discussions and the anonymous referee for helpful comments on a first version of the article.

\section{Proof of the main results}

\V {\bf Proof of Theorem \ref{LinearNeumann}}. In the following we often use {\em squaring of operators}, which is not represented by postcomposition with an injective map.  However, there is a $q \in \LD$ with $ q(x) = x^2 $ for all $ x \in [1/2; \infty) $, and if an operator $A$ has positive spectrum, then $A^2 = q(A)$. Any linear Neumann map is unit-preserving: For $\mathbf{1}$ being the constant unit observable and $ \phi \in \LD$ with $\phi (\R) \subset (1/2, \infty) $ and $ \phi (1) =1 $ we have $ Q(\mathbf{1}) = Q(\phi \o \mathbf{1}) = \phi (Q(\mathbf{1}))$, thus the spectrum of $Q(\mathbf{1})$ is positive, and $Q(\mathbf{1}) = Q(q \o \mathbf{1}) = Q(\mathbf{1}) \o Q(\mathbf{1}) $, thus $Q(\mathbf{1})$ is a projection, which together with the positive spectrum implies $Q(\mathbf{1}) = I_H$. Now we pick two observables $a,b \in G$ whose quantizations $ Q(a_0) =: A_0$ and $ Q(b_0) =:B_0$ do not commute. First of all, defining $ g \in \LD$ by $g := \arctan + \pi$ we replace $a_0$ with $a:= g \o a_0$ and $b_0$ with $ b:= g \o b_0$, obtaining two operators $A:=Q(a)$ and $B:= Q(b)$ with spectrum in $(\pi/2; \infty)$. We still have $[A,B] \neq 0$: Recall the classical von Neumann's theorem\footnote{The original reference for that therorem is \cite{vN1930}, Theorem 10, which is restricted to the case of a separable Hilbert space; the proof however goes through in the general case. For a modern account and the general statement see for example \cite{Rickart1960}, A.2.1.} on the generating operator stating that if $K$ is a set of self-adjoint operators on a Hilbert space that commute with each other, there is a self-adjoint operator $S$ such that for all $ k \in K$ there is $f_k \in \Meas (\R) $ with $k = f_k(S)$ (here, for a measure space $X$, the set $\Meas (X) $ is the set of measurable functions on $X$). Now, assuming that $a $ commutes with $b$, we apply the von Neumann's theorem to $K= \{ a , b \} $, then  there are measurable maps $f_a, f_b$ with $a= f_a (S) $ and $b= f_b(S)$. Taking into account that $g(a_0) = a$, $g(b_0) = b$, for a left inverse $g^{-1} \in \LD$ of $g$ we get $a_0 = g^{-1} (f_1 (S)) $ and $b_0 = g^{-1} (f_2(S))$, so $a_0$ commutes with $b_0$ in contradiction with the assumption. \footnote{The statement also follows e.g. from the useful  formulas in \cite{mR} for commutators with functions of operators that in turn follow from the Helffer-Sj\"ostrand formula.} For the following lemma, we call $x \in G$ {\bf positive} iff there is $ \phi \in \LD $ such that the closure of $ \phi (\R ) $ in $\R$ is contained in $(0; \infty) $, and for two linear endomorphisms $E,F$ in the Hilbert space $H$ we define 

\begin{equation*}
\label{Soll}
S (E,F):=  (EF+FE )^2 - 2 (E^2 F^2 +F^2E^2).
\end{equation*}

\begin{Lemma} 
For any pair of positive $j,l \in G$, we have $S(Q(j), Q(l)) = 0$.
\end{Lemma}

\V {\bf Proof of the lemma}. Elementary arithmetics reveal that

\begin{equation*}
\big( \frac{(j+l)^2 - j^2 - l^2}{2} \big)^2 = (jl)^2 = j^2 l^2 = \frac{(j^2 + l^2 )^2 - j^4 - l^4}{2} ,
\end{equation*}

\V thus if we apply to both sides the linearity of $Q$ and the von Neumann rule applied to $q$ (as above, positivity of $j$ and $l$ allows for an application of the Neumann rule for squaring of $j$, $l$ and $j+l$), so for $J:= Q(j), L:= Q(l)$ we obtain $ \frac{1}{4} (JL+LJ )^2 = \frac{1}{2} (J^2 L^2 +L^2J^2)$ , so we have $S(J,L ) =0$, which concludes the proof of the lemma. \hfill (\qed)

\bigskip

\V Let us denote $A:= Q(a), B:= Q(b)$. To get an idea of the proof of the theorem's general case, let us first assume the existence of a Hilbert basis of  eigenvectors\footnote{But let us recall that there are bounded self-adjoint operators without any eigenvalues, e.g. the multiplication with the identity $x \mapsto x$ in $L^2([0;1])$} of $B$. With the above assumption of an eigenbasis and taking into account that $[A,B] \neq 0$, there is an eigenvector $v$ of $B$ to the eigenvalue $\l$ such that $Av \notin \ker (B - \lambda)$. Then we calculate, using self-adjointness of $A$ and $B$ and writing $w:= Av$ and $S:= S(A,B)$, 

$$ \langle S v,v \rangle=\langle Bw,Bw \rangle + 2 \l \langle Bw,w \rangle - 3 \lambda^2 \langle w,w \rangle  ,$$

\V and this can be made nonzero by replacing $B$ with $\phi(B)$ for $\phi: \R \rightarrow \R$ with $\phi (\lambda) = \lambda$, which does not change $w$ or $\lambda$ in the calculation above.

\V In the general case, let $U \in \mathcal{B} (\R)$, where the latter is the set of Borel subsets of $\R$. We examine $\vert P \o S \o P \vert$ for $P := \mu_B(U) $ (the $B$-spectral measure of $U$). \footnote{Parallelling more closely the proof above by considering $\sup \{ \langle (P \o S_t \o P) (v) , v \rangle : v \in H, \vert v \vert = 1 \} $ instead of $\vert PS_tP \vert$ yields a slightly more complicated proof.} We have $P \o B = P \o B \o P = B \o P  $. Let $B_t:= (Id_\R + t \cdot \chi_{\R \setminus U} )B$, then $B_0=B$ and $ P \o B_t  = P \o B_t \o P  = B_t \o P  $ for {\em all} $t \in \R$. As above, we get $S_t:= S(A,B_t) =0 $, but self-adjointness of $B_t$, $A$, $P$ imply

\bean
&&\vert P S_t  P \vert \\
&=& \big\vert PAB_tAB_tP + PB_tA^2B_tP + PAB_t B_tAP + PB_tAB_tAP - 2PA^2B_t^2 P - 2 PB_t^2A^2P \big\vert\\
&=& \big\vert PAB_tAPB_t + B_tPAAPB_t + PAB_tB_tAP + B_tPAB_tAP - 2 PA^2 PB_t^2 - 2 B_t^2 PA^2P \big\vert \\
&\geq&  \underbrace{\vert PAB_tB_tAP \vert}_{= \vert B_tAP \vert^2 } +2  \vert B_tAP \vert \cdot \vert \underbrace{B_tPA}_{=BPA} \vert  - 3 \vert \underbrace{B_tPA}_{=BPA} \vert^2 \rightarrow_{t \rightarrow \infty} \infty
\eean

\V where we use the formulas $\vert W \vert = \vert W^\dagger  \vert $, $ \vert W W^\dagger \vert = \vert W \vert^2$, $\vert B_tPA \vert = \vert BPA \vert $. Finally, $ \lim_{t \rightarrow \infty} \vert B_tAP \vert = \infty$, as there is some $U \in \mathcal{B}(\R)$ with $P^\perp A P \neq 0$ for $P:= \mu_B(U)$: Assume the opposite, then due to self-adjointness we have $P_U^\perp A P_U = 0 = P_U A P_U^\perp$, so $[P_U, A] = [P_U, (P_U + P_U^\perp) A (P_U + P_U^\perp)] = 0$ for all $ U  \in \mathcal{B} (\R)$. Thus $[A,B] =0$ as $B = \int_\R I_\R (x) d \mu_B (x) $. \hfill \qed

\bigskip

\V {\bf Proof of Theorem \ref{Peetre}.} As hypothesis and conclusion of the the theorem are invariant under composition with trivializations (being local diffeomorphisms), it suffices to show the statement for $M$ an open subset of a Hilbert space $Z$ and trivial vector bundles of fibers $V$ resp. $W$. Let us, for $x \in M$, denote by $N_x$ the set of open neighborhoods of $x$. 

\begin{Lemma}
\label{Lemma1}
Assume the hypothesis of the theorem, then:

\bean
\forall x \in M \forall C>0 \exists U \in N_x \exists k \in \N \forall y \in U \setminus \{ x \} \forall s \in C^\infty (U,V):\\ (j^k s) (y) = 0 \Rightarrow \vert Ls (y) \vert < C .
\eean 
\end{Lemma}

\V {\bf Proof of the Lemma.} Assume the opposite, then there is a sequence $y \in M^\N$ in $M$ with $\lim_{n \rightarrow \infty} (y_n) = x$ and a sequence $r \in (0; \infty)^\N $ of radii such that, for $B_k := B(y_k, r_k)$, we have $cl(B_k) \cap cl(B_l) = \emptyset \forall k \neq l$, and there are $s_k \in C^\infty (M, V)$ with $(j^k s_k) (y_k) = 0 $ and $ \vert L s_k (y_k) \vert \geq C >0 $. We want to produce a contradiction by evaluating separately at the even and at the odd points the image under the operator of a carefully chosen section. Let $a \in C^\infty (Z, [0;1])$ with $a(B(0,1/2)) = \{ 1 \}$ and $a(Z \setminus B(0,1) ) = \{ 0 \}$ with $\sum_{j=0}^k \sup \{ \vert d^j a (x) \vert : x \in Z \} =: E_k < \infty$; such an $a$ can easily be constructed, chosen radially invariant. For all $k \in \N$ we have $(j^{2k} s_{2k} ) (y_{2k}) =0$, and the mean value theorem applied to $ \vert d^j s_{2k} \vert \o c$ for a radial curve $c$ implies that there is $\rho_{2k} \in (0;r_{2k})$ such that for all $\d \in (0;\rho_{2k})$ we have 

$$ \sum_{\vert j \vert <k} \sup \{ \vert d^j s_{2k} (y) \vert : y \in B(y_{2k}, \d \} \leq \frac{1}{M_k} (\frac{\d}{2})^k  $$

\V With $a_{2k, \d}: Z \rightarrow [0;1], a_{2k, \d} (z) := a(\frac{z-y_{2k}}{\d})$ we get

$$ \max_{j \leq k } \sup \{ \vert d^j (a_{2k} s_{2k}) \vert : y \in B(y_{2k}, \d) \} \leq 2^{-k}) .  $$

\V By comparison with the geometric series and uniform convergence we see that $q: z \mapsto \sum_{k=0}^\infty a_{2k} (z) \cdot s_{2k} (z) $ is a smooth function from $Z$ to $V$. As

$$ s_{2k} \vert_{B(y_{2k}, \d/2)}  = a_{2k, \d} \cdot s_{2k} \vert_{B(y_{2k}, \d/2))},$$

\V we get $ \lim_{k \rightarrow \infty} \vert L q (y_{2k}) \vert \geq C$, and continuity of $Lq$ implies 

\bea
\label{even}
\vert Lq(x) \vert \geq C >0.
\eea

\V On the other hand, tracing the odd points we get $Lq(y_{2k+1}) =0$ as $q \vert_{B_{2k+1}} =0 $ and $\supp Lq \subset \supp q \subset Z \setminus B_{2k+1} $. Continuity of $Lq$ implies $Lq (x) = 0$, in contradiction to Eq. \ref{even}. \hfill (\qed)

\begin{Lemma}
\label{Lemma2}
Assume the hypothesis of the theorem, then:

\bean
\forall x \in M  \exists U \in N_x \exists k \in \N \forall y \in U  \forall s \in C^\infty (U,V):\\ (j^k s) (y) = 0 \Rightarrow  Ls (y)  =0 .
\eean 
\end{Lemma}

\V {\bf Proof of the Lemma:} Fix $x \in M$ and $C>0$, then there are $U$ and $k$ as in Lemma \ref{Lemma1}. Assume that there is a $y\in U \setminus \{ x \}$ with $j^k s(y) = 0$ and $ \vert Ls (y) \vert  = b >0$. Then consider $ \tilde{s} := \frac{2C}{b} \cdot s \in C^\infty (U,V)  $, then $ j^k \tilde{s} (y) =0 $ and $\vert L \tilde{s} (y) \vert = 2C >C$, in contradiction to Lemma \ref{Lemma1}. Finally, $Ls(x) = 0$ holds by continuity of $Ls$. \hfill  (\qed)

\bigskip

\V {\bf Proof of the theorem, ctd.:} Now, for $U,k$ as in Lemma \ref{Lemma2}, $y \in U$ and $b \in J^k \pi_y $, there is a map $s \in C^\infty (U,V)$ with $b = j^k s (y)$, and we define $u((j^k s) (y) ) := L s(y)$, which is well-defined due to Lemma \ref{Lemma2}. \hfill \qed

\bigskip

\V {\bf Proof of Theorem \ref{LocalNeumann}:} Let $f \in G$, then Peetre's Theorem above implies that in a small neighborhood $U$, $Q(f)$ is a differential operator of, say, order $k$. As the order of a differential operator is multiplicative under taking powers, $Q(\sqrt[k+1] (f)) $ is a $(k+1)$-th root of $Q(f)$ and so cannot be a differential operator, not even in a smaller neighborhood, contradiction. \hfill \qed

\bigskip

\V With the arguments above for an exact validity of the von Neumann rule, it appears worthwhile to look for nonlinear quantization maps\footnote{Another approach is treating bosonic degrees of freedom structured by commutators as secondary, emergent objects and only fermionic degrees of freedom displaying anticommutators as truly fundamental (a possible limitation of this approach is the result in \cite{aK} for finite-dimensional systems). For a non-quantization version of this idea, see \cite{fF}.}, e.g. in the spirit of the proposals of Kibble \cite{Kibble} and Weinberg \cite{Weinberg} (see also \cite{Polchinski}, \cite{Jordan}). However, in those approaches not only $Q$ is nonlinear, but also the $Q(f) $ are, and there does not seem to be a good suggestion for how to replace the Born rule in this context. Interestingly, already Wigner \cite{Wigner} concluded from a gedanken experiment (in a certain double sense) that quantum theory cannot be linear, independently of the von Neumann property.

%
%
%
%
%
%
%
%
%
%
%
%
%

\section{Conclusion}

The upshot of the considerations above is that any Neumannian map (and thus every reasonable quantization map) is neither linear nor local. 

\V Of course, considering this somehow awkward finding, one can ask whether quantization is the right approach at all. Specifically, the concept of quantization, despite its success in the standard model of particle physics, is sometimes subjected to the criticism that a truly fundamental structure should rather be a map in the reverse direction. This goes under the name 'dequantization'. Even in several quantization schemes, inverses of the respective quantization map play a certain role, e.g. the Wigner transform in Weyl quantization (\cite{HDS}, \cite{Case}) and the Berezin symbol in Berezin-Toeplitz quantization (\cite{mE}, \cite{mS}) (note that in Geometric Quantization, a simple computation shows that for the quantization map $Q$ of Geometric Quantization and for $\s$ being the principal symbol of a differential operator, we get $\s \o Q (f) = {\rm sgrad} (f)$, the symplectic gradient of $f$). However, inverting the direction of quantization or, more generally, allowing for quantization {\em relations} instead of quantization maps, would make a difference only if there were two measurement devices 'measuring the same classical quantity' (in the classical decoherence regime) but could be represented by two different operators in a Hilbert space in a systematical way. Whether this is the case seems to be unknown at present (\cite{Embacher}). The fundamental importance of the canonical commutation relation seem to indicate the opposite, suggesting that the commutator of {\em every} measurement apparatus associated to the classical momentum and {\em every} measurement apparatus associated to the classical position should be a multiple of the identity. If two momentum measurement devices yield identical results in the classical decoherence regime but are represented by two different operators $P$ and $\tilde{P}$, assume that $[P, \tilde{P}] = 0$ and $[P, X] = c I_H = [\tilde{P}, X ] $. Then $R:= \arctan (P - \tilde{P})$ is bounded and $[R,X ] = 0 = [R,P]$, and if a relational Neumann property holds (stating that for any apparatus with quantum operator $A$ related to an observable $f$ there is an apparatus with  quantum operator $\phi(A)$  related to the classical quantity $\phi \o f $), then $R$ commutes with a family of operators related to {\em every classical quantity} (by well-known Weierstra\ss -like theorems, see e.g. \cite{BGG}), which by the usual assumption of irreducibility means that $R$ is a constant, i.e. $ \tilde{P} = P + k I_H$, in contradiction to the fact that $P$ and $\tilde{P}$ coincide classically. Thus {\em if there is a dequantization theory yielding more correct predictions than quantization, one should be able to find two momentum measurement devices either not commuting with each other or at least one of which does not have commutator $c I_H$ with position.} 

\bigskip 
 
\V A last comment concerning the experimental accessibility of nonlinear observables: It is easy to prove that any linear Neumannian quantization map $Q$ satisfies $Q(a \cdot b) = \frac{1}{2} (Q(a) \o Q(b) + Q(b) \o Q(a))$ for any two observables $a,b$. Thus linearity of quantization can in principle be tested by analyzing the effect of devices measuring $x \cdot p$ for a point particle, which can be realized e.g. by examining interference patterns on a screen perpendicular to a constant magnetic field $B$ and a appropriately coherent beam of nonrelativistic charged particles parallel to the screen. Due to the Lorentz force, the distance of the classical hit point on the screen from the source is $\sqrt{x p}$ for $x$ being the initial distance of the particle from the screen and $p$ is its momentum perpendicular to $B$.

\end{document}